\documentclass[aps,pra,amssymb,amsmath,reprint,noshowpacs,superscriptaddress,floatfix]{revtex4-1}

\usepackage{graphicx}
\usepackage{hyperref}
\usepackage{siunitx}
\usepackage[capitalize]{cleveref}
\usepackage[usenames]{color}
\usepackage[normalem]{ulem}
\usepackage{soul}
\usepackage[disable,textsize=scriptsize]{todonotes}

\newcommand{\tens}{\stackrel{\leftrightarrow}}
\newcommand{\im}{\mathrm{i}}
\newcommand{\kk}{\kappa}
\newcommand{\oo}{\omega}

\begin{document}

\title{Controlling nanoantenna polarizability through backaction via a single cavity mode}
\author{Freek Ruesink}
\thanks{These authors contributed equally}
\affiliation{Center for Nanophotonics, AMOLF,
Science Park 104, 1098 XG Amsterdam, The Netherlands}
\author{Hugo M. Doeleman}
\thanks{These authors contributed equally}
\affiliation{Center for Nanophotonics, AMOLF,
Science Park 104, 1098 XG Amsterdam, The Netherlands}
\affiliation{Van der Waals-Zeeman Institute, University of Amsterdam, 
Science Park 904, PO Box 94485, 1090 GL Amsterdam, The Netherlands}
\author{Ewold Verhagen}
\affiliation{Center for Nanophotonics, AMOLF,
Science Park 104, 1098 XG Amsterdam, The Netherlands}
\author{A. Femius Koenderink}
\affiliation{Center for Nanophotonics, AMOLF,
Science Park 104, 1098 XG Amsterdam, The Netherlands}
\affiliation{Van der Waals-Zeeman Institute, University of Amsterdam, 
Science Park 904, PO Box 94485, 1090 GL Amsterdam, The Netherlands}
\email{f.koenderink@amolf.nl}

\begin{abstract}
The polarizability $\alpha$ determines the absorption, extinction and scattering by  small particles. Beyond being  purely set by scatterer size and material, in fact polarizability can be affected by backaction: the influence of the photonic environment on the scatterer.
As such, controlling the strength of backaction provides a tool to tailor the (radiative) properties of nanoparticles. 
Here, we control the backaction between broadband scatterers and a single mode of a high-quality cavity. 
We demonstrate that backaction from a microtoroid ring resonator significantly alters the polarizability of an array of nanorods: the polarizability is renormalized as fields scattered from -- and returning to -- the nanorods via the ring resonator depolarize the rods. 
Moreover, we show that it is possible to control the strength of the backaction by exploiting the diffractive properties of the array. 
This  perturbation of a strong scatterer by a nearby cavity has important implications for hybrid plasmonic-photonic resonators and the understanding of coupled optical resonators in general.
\end{abstract}
\date\today

\maketitle

\begin{figure}
\includegraphics[]{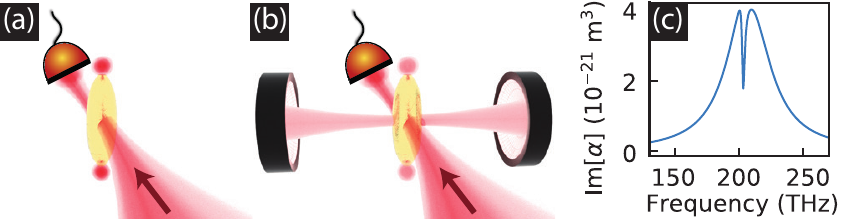}
\caption{
(a) Single polarizable scatterer.
(b) A simple Fabry-P\'erot cavity modifies the Local Density of Optical States and alters the scattering properties of a plasmonic scatter.
(c) A spectrally narrow cavity mode can suppress the imaginary part of the polarizability of a plasmonic scatterer.
}
\label{fig:concept}
\end{figure}

The scattering, absorption and extinction cross-section of small scatterers is often attributed to the dielectric properties of the particle, \textit{i.e.,} the scatterer's volume, shape and its refractive index with respect to the host medium~\cite{Bohren1983}.
Central to this argument, for scatterers with a physical size much smaller than the wavelength, is the so-called \textit{polarizability}, which contains the frequency-dependent susceptibility that quantifies the strength of the dipole moment induced in the scatterer by an incident field.  
A rather subtle notion is that the polarizability also depends on the mode structure offered by the photonic environment~(\cref{fig:concept}). 
To illustrate this, consider that  extinction, \textit{i.e.}, the total power that a scatterer extracts from an incident beam~\cite{Bohren1983} is directly proportional to the imaginary part of polarizability. 
According to the optical theorem~\cite{DeVries1998}, this power is distributed over Ohmic heating and scattering, with the contribution of scattering being  proportional to the squared magnitude of polarizability and the Local Density of Optical States (LDOS)~\cite{Novotny2012}. 
The fact that LDOS, \textit{i.e.}, the number of available photonic modes for the scatterer to radiate into, enters the polarizability  is known as backaction: a  correction on the total field that drives a polarizable scatterer.
This correction is neglected in standard (Rayleigh) scattering theory~\cite{Bohren1983}. However, even for a single scatterer placed in free-space, backaction leads to additional damping (depolarization) and thus needs to be integrated in a self-consistent description of any system~\cite{DeVries1998,Agio2013}.    
Although backaction effects on quantum emitters~\cite{Drexhage1970} have been routinely studied, the only experiment in which backaction on a plasmonic scatterer was rigorously evidenced was performed by Buchler et al.~\cite{Buchler2005}.
They showed that the spectral width of the plasmon resonance of a nanoantenna can be modulated  when the nanoantenna approaches a reflector.
However, to our knowledge no experiments have manipulated the \textit{magnitude} of the polarizability as a function of the bath of photonic modes to which it couples, despite the fact that the polarizability of nanoantennas directly controls their effectiveness in their main application areas, \textit{i.e.}, near-field enhancement and light-matter interaction strength. 

\begin{figure}
\includegraphics[]{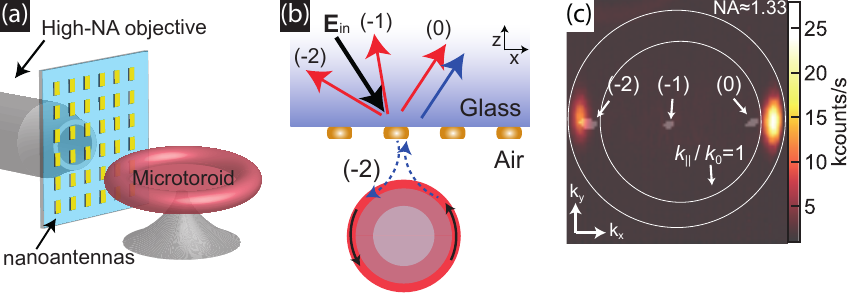}
\caption{ 
(a) Cartoon of the hybrid cavity-antenna system.
(b) For some incoming field $\mathbf{E}_{\mathrm{in}}$, the ${(-2)}$ diffraction order associated with the array evanescently couples to the toroid. 
Backaction from the cavity on the array is measured in the specular reflection signal. 
We block the ${(-2)}$ and ${(-1)}$ diffraction orders back into the glass, as well as direct radiation of the cavity into the glass substrate.
(c) An overlay of Fourier-images obtained via back-focal-plane imaging. 
The transparent white blobs indicated by arrows are diffraction orders.
The $(-2)$ order overlaps with one of the cavity modes (indicated with the color scale), exciting the cavity mode \textit{via} the antenna-array. 
}
\label{fig:intro}
\end{figure}

Here we experimentally investigate backaction on polarizability in a hybrid cavity-antenna system~(\cref{fig:intro}a),  demonstrating a strongly modified extinction response of an array of gold nanorods due to backaction imparted by a single whispering-gallery-mode (WGM) of a microtoroid ring resonator. 
At conditions where the cavity offers a high mode density for the scatterers to radiate into, the   nanorods' susceptibility to an incoming field is suppressed: the cavity mode density thus effectively depolarizes the nanorods~(\cref{fig:concept}c), yielding an experimental signature that relates to electromagnetically induced transparency~\cite{Boller1991}.
Our experiments reveal that it is possible to control the strength of the measured backaction by careful tuning of a diffraction order of the array, phase-matching its wavevector with the WGM of the cavity.
Using a coupled-oscillator model we retrieve an antenna-cavity cooperativity and provide a lower bound on the cavity Purcell factor~\cite{Purcell1946} at the lattice origin.
Our results have large relevance in the context of recent proposals on hybrid plasmonic-photonic resonators~\cite{Ameling2010,Barth2010,Ahn2012,Frimmer2012,Schmidt2012,Xiao2012,Ameling2013,Foreman2013,Doeleman2016,Gurlek2017} as a unique venue for huge Purcell factors~\cite{Purcell1946} and quantum strong coupling with single emitters. 
While the most intuitive consideration for such a proposal is to assess how scatterers perturb  cavity resonances~\cite{Ruesink2015}, in fact, this work shows that one rather has to ask what opportunities the cavity offers to control antenna polarizability.

An ideal experiment to probe  cavity-induced backaction would directly measure the complex-valued polarizability $\alpha$ of a scatterer in  presence and absence of the microtoroid.
This is not a trivial task: polarizability is not a directly measurable quantity in optics. 
Instead one has to rely on far-field measurements of extinction and scattering cross sections to  deduce  $\text{Im}[\alpha]$ and $|\alpha|^2$ respectively.  
Such quantitative polarizability measurements on single nano-objects are challenging even for single scatterers in an uniform environment~\cite{Husnik2012,Husnik2013}.
The proximity of the cavity further complicates the task of strictly probing the scatterers only.  
Practically, this means that direct excitation of the cavity mode by the incident beam, as well as radiation from the cavity directly into the detection channel, should be prevented,  as both would contaminate the interrogation of the scatterer's response.
We approach these constraints by a combination of experimental techniques. 
First, we use a WGM resonator that only allows in-~and~outcoupling of light under select wavevector matching conditions. 
Second, we use an array of antennas, as opposed to a single antenna, to obtain a strong extinction-like signal that can be probed in specular reflection with a nearly collimated plane wave, again using wavevector conservation to separate the extinction channel from all other scattering channels. 

We use gold nanorods with length(width) of $\SI{400}{nm}(\SI{120}{nm})$ and thickness of $\SI{20}{nm}$ placed on a glass substrate in an array with \SI{800(1500)}{nm} pitch along the long(short) axis of the rods.
In absence of the cavity, the array exhibits a broadband resonant response centered at $\omega_\mathrm{a}/2\pi\approx\SI{208}{\tera\hertz}$~\cite{SM,Ruesink2015} (linewidth $\gamma\approx\SI{55}{THz}$), while the microtoroid~\cite{Armani2003} (linewidth $\kappa\approx\SI{30}{MHz}$) is resonant at slightly red-detuned frequency $\omega_\mathrm{c}/2\pi\approx\SI{194.4}{\tera\hertz}$.
The incident drive field is polarized (s-polarization) along the principal dipole axis of the rods, which themselves are oriented to match a high-Q TE-polarized mode of the microtoroid. 
The response measurements on the array involve a high-NA objective ($\text{NA}\approx1.33$, used with index-matching oil) operated in reflection. 
Focusing the incoming laser beam onto the back-focal-plane (BFP) of the objective gives  precise control over the angle of incidence of the drive field.
For scatterers arranged in a periodic array, scattering takes the form of diffraction into well-defined angles (wavevectors,~\cref{fig:intro}b). 
We discard the ${(-2)}$ and ${(-1)}$ diffraction orders propagating back into the substrate using Fourier-filtering such that our detector is only sensitive to the specular reflection signal.
In addition we employ a real-space filter, selecting a circular area of $\SI{\sim 4.5}{\micro\meter}$ in diameter, to reduce background signals not originating from antennas coupled to the cavity.
To illustrate this experimental arrangement, \cref{fig:intro}c displays an overlay of Fourier-space data obtained by BFP imaging (without Fourier-filter).
We identify 1) the radiation profile of the two propagating cavity modes, obtained by direct excitation of the cavity using an evanescently coupled tapered fiber (color scale), and 2) the position of the three diffraction orders of the array (indicated by arrows). 
The incoming wavevector ($k_\parallel/k_0=0.84$) is chosen such that the $(-2)$ diffraction order of the array overlaps with one of the propagating whispering gallery modes in the microtoroid, allowing the incoming field to efficiently scatter to the cavity mode \textit{via} the antennas.
Our system thus allows for a proper backaction measurement:  the antennas can couple to the cavity, yet the detected signal is exclusively a probe of antenna polarizability.
Any change in detected signal upon approaching the cavity can thus be directly attributed to cavity-mediated backaction fields that renormalize the antennas response.

\begin{figure}
\includegraphics[]{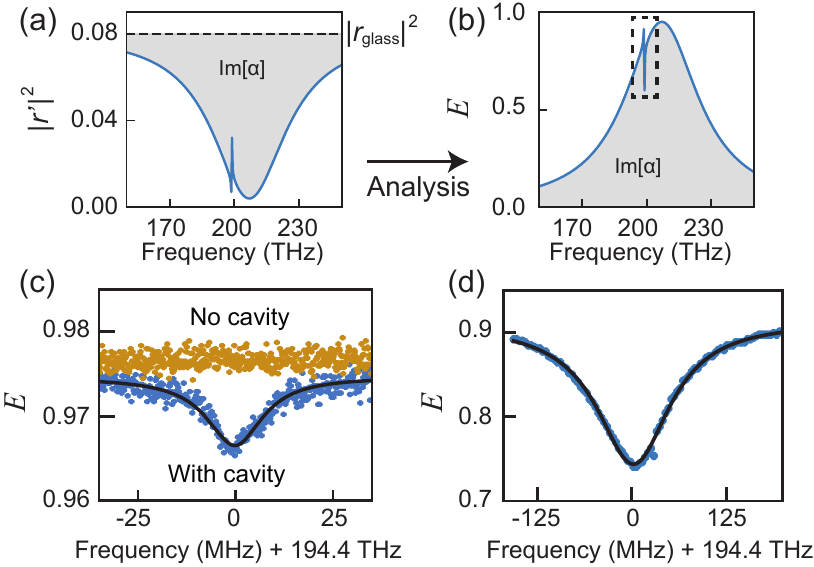}
\caption{
(a) Sketch of a typical reflectance signal $|r'|^2$ as measured in the experiment. The plasmon feature introduces a broadband dip, with respect to the non-resonant $|r_\text{glass}|^2$ value. 
(b) Sketch of the extinction $E$, obtained from (a). 
We expect the cavity mode to reduce the antenna-extinction (and thus $E$) in a narrow frequency band. 
Dashed box: experimentally accessible frequency regime.
(c),(d): Experiment. (c) With the cavity present (blue points), the extinction $E$ decreases  by $1\%$ at the cavity frequency, a feature that is absent without cavity (orange points).
(d) At smaller cavity-array distance the dip increases to $25\%$, indicating a strong suppression of antenna extinction.
The cavity linewidth increases compared to (c) as a result of increased cavity losses. 
}
\label{fig:analysis}
\end{figure}

In the experiment we probe the antennas through zero-order reflectance at a small (around 30 degree) incident angle, where zero-order reflectance is a direct measure of extinction, \textit{i.e.}, $\text{Im}[\alpha]$~\cite{Hulst1957}.
Since extinction is usually associated to zero-order transmittance and not reflectance, this claim requires substantiation.  
The antennas lie on a glass-air interface which in itself is reflective. 
To predict the lineshape in reflectance, we quote an expression from De Abajo~\cite{DeAbajo2007} for the near-normal incidence specular reflection $r'$ of  particle arrays
\begin{equation}
r' = r_\text{glass} +\frac{ 2\pi \im k \alpha}{A}, 
\end{equation}
where $A$ is the unit-cell area of the array, $k$ the wavenumber and $\alpha$ the antenna polarizability~\footnote{Note that eq.(8) in~\cite{DeAbajo2007} uses a lattice-normalized polarizability.}.
Importantly, one expects a reflection baseline given by the glass-air interface (non-resonant, real-valued $r_\text{glass}$) together with a broadband plasmon feature.  
One can show \cite{SM} that in our system the plasmon feature leads to a broadband reflectance \textit{minimum} that  primarily reports on $\text{Im}[\alpha]$ (\cref{fig:analysis}a).  
In essence, destructive interference causes a reduction in reflectance, similar to the textbook scenario of extinction measurements that measure destructive interference between forward scattered light and the direct beam. 
In analogy to standard transmittance measurements probing extinction, we here identify the extinction $E$ via $E\equiv1-|r|^2$, with the normalized reflectance $|r|^2$ given by $|r|^2\equiv|r'|^2/|r_\mathrm{glass}|^2$.
The use of $|r|^2$ has the advantage that results obtained at different excitation angles (leading to different values of $r_\mathrm{glass}$) are more easily compared.
Moreover, the introduction of the variable $E$ simplifies the interpretation of our experiment: a decrease in antenna-extinction (increasing $|r|^2$) is mapped to decreasing values for $E$.
Our prediction is that the polarizability will show a reduction over a narrow spectral region~\cite{Frimmer2012,Doeleman2016}, which will hence also appear as a minimum in $E$ (\cref{fig:analysis}b),  once the antennas are subject to backaction through the cavity mode, i.e., once they are offered the additional possibility of radiation damping due to the Purcell factor associated with the cavity mode.

\Cref{fig:analysis}c displays the response of the antenna array in absence (orange points) and presence (blue points) of the cavity for an incident beam with $k_\parallel/k_0=0.84$.  
The narrow frequency window displayed in \cref{fig:analysis}c is close to the plasmon resonance.
This is evident from the fact that $E$ is close to unity, meaning that $|r|^2$ is close to zero. 
Comparing the trace without cavity and with the cavity approached to several microns distance away (antennas weakly couple to the cavity) shows a small backaction effect of the cavity on the array, visible as a $\sim1\%$ dip in $E$.
This dip is tantamount to a \textit{reduction} in the extinction that the antennas cause when they are offered the cavity as an additional channel to radiate into.
Expressed in polarizability, our measurement implies a change in $\text{Im}[\alpha]$ due to backaction,  occurring over a narrow bandwidth that is commensurate with the linewidth of the high-Q cavity mode. 
In \cref{fig:analysis}c the cavity-array distance was several microns, limiting the backaction experienced by the antennas.
Moving the cavity closer to the array results in much larger effects.  
For instance, \cref{fig:analysis}d shows a $>25\%$ change in polarizability when approaching the cavity to within approximately 1 micron (about 4 times the evanescent decay length of the mode) from the antennas.   
This is direct evidence that the \emph{magnitude} of polarizability can be substantially controlled by the photonic environment. 

While our experiment probes several antennas, it was previously realized that for single antennas  the polarizability modification must be directly linked to the cavity Purcell factor at the location of the antenna~\cite{Frimmer2012,Doeleman2016}.
In other words, one viewpoint on our experiment is that it evidences that the polarizability of a nano-antenna is modified, which is  mathematically expressed as $\alpha^{-1}=\alpha_0^{-1}-G$, with  $\text{Im}[G]$ the LDOS and $\text{Re}[G]$ the Lamb shift~\cite{Lamb1947} provided by the cavity mode. 
As such, an antenna is analogous to a quantum emitter in the sense that it probes the LDOS of the cavity.
The effect of an LDOS peak, however, is distinctly different: the antenna emission is quenched on resonance rather than, as would be the case for an emitter, enhanced. 

\begin{figure}
\includegraphics[]{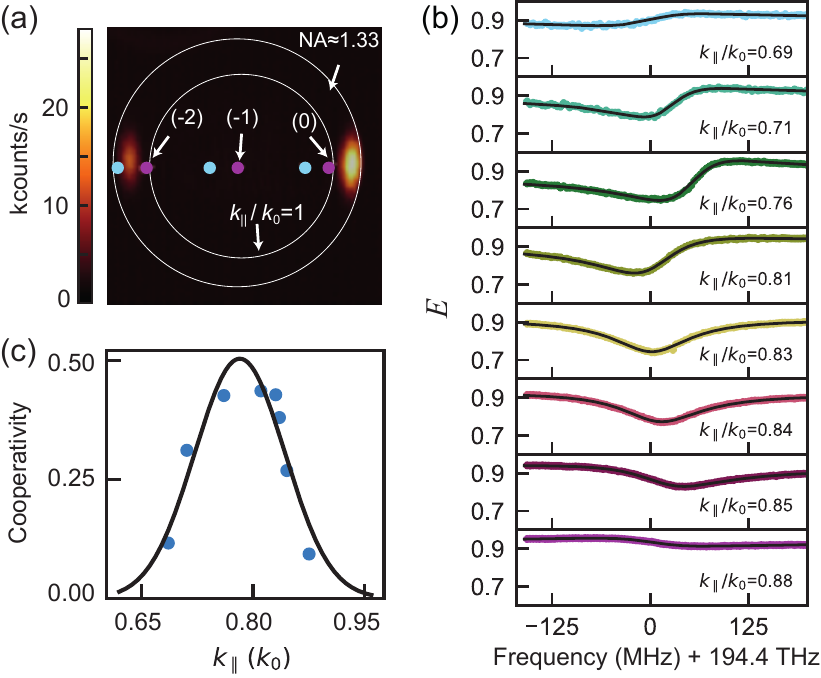}
\caption{
(a) Fourier-image overlay that shows the position of the diffraction orders at the start (blue dot, $k_\parallel=0.69 k_0$) and stop (pink dot, $k_\parallel=0.88 k_0$) values of the $k_\parallel$ sweep displayed in panel (b).
(b) The strength and lineshape of the backaction strongly depend on the incoming angle. The black lines are fits of our coupled-mode model.
(c) Values for the cooperativity obtained from fitting our coupled-mode model to the spectra in panel (b). 
Black line: fit with a Gaussian lineshape. 
}
\label{fig:data}
\end{figure}

The fact that in our experiment the mode density provided by the cavity results from a single Lorentzian mode offers an alternative viewpoint.  
In essence, the reduction of polarizability over the cavity bandwidth can be viewed as a `transparency'  feature in direct analogy to electromagnetically/plasmon/optomechanically induced transparency~\cite{Boller1991,Zhang2008,Liu2009,Weis2010,Safavi-Naeini2011a}.
In these systems, a broad resonator (here: plasmonic scatterer) is rendered `transparent' in its susceptibility to driving over a narrow frequency band due to coupling to  a narrow resonator (here: WGM resonator), even though that narrow resonator is not directly driven.  
Beyond purely Lorentzian transparency dips, one can obtain Fano-type~\cite{Fano1961} line shapes depending on the phase of the coupling constants that connect the broad and narrow resonance. 
Inspired by this analogy we explore the shape of the backaction feature by varying the angle of incidence of the incoming drive field. 
As shown in \cref{fig:data}a, this effectively sweeps the ${(-2)}$ diffraction order over the finite k-space width of the cavity mode, thus varying the degree to which the array and the cavity mode are coupled.
From the resulting spectra (\cref{fig:data}b) we qualitatively observe a dependence of the backaction strength and lineshape on the incoming angle, which is expressed as a varying depth and asymmetry of the cavity-induced dip.
In line with the phase-matching argument, visual inspection of \cref{fig:data}a and \cref{fig:data}b shows that cavity-mediated backaction is most prominent when the cavity mode profile and the $(-2)$ diffraction order of the array experience better overlap. This behavior is verified using analytical coupled dipole calculations \cite{SM}.
 
Full quantification of the backaction is not straightforward, as it requires analysis of the Fano lineshapes.  
A detailed multiple scattering analysis particular for our system~\cite{SM} shows that the plasmon antennas in our experiment are simultaneously subject to the resonant backaction of the cavity and a nonresonant backaction term from the interface on which the antennas are placed (glass-air)~\cite{Novotny2012,Kwadrin2016}. The nonresonant backaction is governed by the complex Fresnel coefficient associated with the interface, which exhibits a phase change for the (evanescent) $(-2)$ diffraction order upon sweeping $k_{||}/k_0$.
In our experiment  we measure the scatterers' response in the presence of all backaction, which is a coherent sum of the broadband interface-induced backaction plus the resonant cavity-mediated backaction. Sweeping $k_{||}$ thus directly affects the Fano lineshape that we observe. 
We develop a simple model based on coupled-mode theory~\cite{Haus1984} that can  disentangle the resonant backaction from the nonresonant background. 
Treating the array and cavity as resonators, coupled at rate $g$, both described by a Lorentzian response with complex field amplitudes $a$ and $c$, respectively, we solve the driven system 
\begin{equation}
\begin{pmatrix}
\Delta_\mathrm{a} + \im \gamma /2 &  g \\
g & \Delta_\mathrm{c} + \im\kk/2
\end{pmatrix}
\begin{pmatrix}
a \\ c
\end{pmatrix}
= 
\begin{pmatrix}
\im \sqrt{\gamma_\mathrm{ex}} s_\mathrm{in} \\ 0
\end{pmatrix}
\end{equation} 
for $a$.
Here we defined $\Delta_\mathrm{a}\equiv\omega-\oo_\text{a}$ and $\Delta_\mathrm{c}\equiv\omega-\oo_\text{c}$, where $\oo$ is the frequency of the incident field $s_\mathrm{in}$ driving the array and $\gamma_\mathrm{ex}$ the rate at which the array and input/output channel are coupled. 
Next, we use the input-output relation $s_\mathrm{out}= s_\mathrm{in} - \sqrt{\gamma_\mathrm{ex}} a$~\cite{Haus1984} (such that $r=s_\mathrm{out}/s_\mathrm{in}$) and parameterize the coupling via the cooperativity $C=4g^2/(\gamma\kappa)$, the determining quantity for the strength of the sharp spectral feature observed in electromagnetically/optomechanically induced transparency~\cite{Aspelmeyer2014,Lodahl2015}. 
We obtain 
\begin{equation}
|r|^2 = \Big|\exp(\im\phi) -  \frac{2\eta}{1 
 + \frac{C}{\frac{2 (\Delta_\mathrm{c}-\Delta)}{\im \kappa}+1 }} \Big|^2,
\label{eq:alpha_fit_R}
\end{equation}
where $\eta\equiv\gamma_\mathrm{ex}/\gamma$, $\Delta$ a parameter capturing small changes in $\oo_\mathrm{c}$, $\phi$ an arbitrary phase pickup and we assumed $\oo\approx\oo_\mathrm{a}$.
We use \cref{eq:alpha_fit_R} to fit our experimental data in \cref{fig:data}b, yielding values for $C$ as a function of $k_\parallel$ (\cref{fig:data}c, blue points). 
A Gaussian lineshape (black line) is fit (center(width):~$k_\parallel/k_0\approx0.78(0.14)$) to the blue data points, giving a maximum cooperativity of $C\approx 0.5$. 
Notably, the width and center of the Gaussian agree with expected values based on a cross-cut of the cavity mode profile observed in \cref{fig:data}a.  
The cooperativity in the limit of a \textit{single} scatterer and single cavity mode, is directly equivalent to the product of the scatterer albedo $(A)$ and the cavity Purcell factor $(F)$ at the location of the scatterer~\cite{SM}.
In our experiment the cooperativity can not be directly cast into a Purcell factor, as we probe  an array of antennas at specific wavevector,  meaning that we probe a lattice-sum dressed polarizability (see de Abajo~\cite{DeAbajo2007}) that experiences backaction from a wavevector resolved mode density. 
Using calculations on a lattice of scatterers~\cite{SM}, we estimate that the measured cooperativity of $C=0.5$ actually corresponds to a value of $C=1.7$ as it is felt by a single antenna, without a lattice, located at the lattice origin. 
Considering that $A<1$, the backaction feature in our experiment is tantamount to a modest Purcell factor of $F\geq1.7$.  
Obviously this effect could be much stronger in experiments where the scatterers are placed right in the mode maximum, as opposed to the arrangement in our set up where scatterers are placed at in the evanescent tail (estimated decay length of \SI{230}{nm}~\footnote{Using eq.~(1.74) in~\cite{Anetsberger2010} for $n=1.46$}) of the cavity mode at approximately $\SI{1}{\micro\meter}$ distance.

Concluding, we have shown that cavity backaction can alter the polarizability of an array of scatterers, and that the strength of the backaction can be controlled via the incoming drive field. 
Whereas in this work the Purcell enhancement provided by the cavity effectively depolarizes the nanorods, which is related to the fact that the cavity and array are nearly resonant, it has been predicted that both an increase and decrease in polarizability can be obtained by controlling the detuning between cavity and scatterers~\cite{Doeleman2016}. This type of control is instrumental for exploration of the field of hybrid cavity-antenna systems that promises to combine plasmonic field enhancements derived from scatterers with microcavity Q's with advantages for single-photon sources, strong coupling to single quantum emitters, as well as classical applications like single-molecule sensing~\cite{Xiao2012,Frimmer2012,Vollmer2012,Doeleman2016,Gurlek2017}.
Our results show the feasibility of such an approach. 

We thank Martin Frimmer for stimulating discussions during the start of this project. 
This work is part of the research program of the Netherlands Organization for Scientific Research (NWO). 
AFK acknowledges an NWO Vici grant.

%\bibliography{alpha_biblio,suppl}

%

% --------------------------

% ----------- SI

% ---------------------------
\newpage

\newcommand{\oa}{\omega_\text{a}}
\newcommand{\rr}{\mathbf{r}}
\renewcommand{\oc}{\omega_\text{c}}

\renewcommand{\theequation}{S\arabic{equation}}
\renewcommand{\thefigure}{S\arabic{figure}}

\setcounter{figure}{0}
\setcounter{equation}{0}

%%%%%%%%%%%%%

%%New Commands
%figure width and comment out
\newcommand{\figwidth}{0.4\linewidth} %% specifies size for pictures
\newcommand{\commentOut}[1]{}
%to do
%\newcommand{\ToDo}[1]{\textbf{\textcolor{blue}{[#1]}}}
%\newcommand{\todo}[1]{\ToDo{#1}}
% frequencies omega
\newcommand{\ooc}{\omega_\mathrm{c}}
\newcommand{\oz}{\omega_0}
\newcommand{\OO}{\Omega}
\newcommand{\doo}{\delta \oo}
\newcommand{\doc}{\delta \ooc}
% damping rates kappa/gamma
\newcommand{\dkk}{\delta \kk}
\newcommand{\ki}{\kappa_\mathrm{i}}
\newcommand{\kex}{\kappa_\mathrm{ex}}
\newcommand{\kexp}{\kappa_\mathrm{ex}^{\mathrm{p}}}
\newcommand{\kr}{\kappa_\mathrm{r}}
\newcommand{\gi}{\gamma_\mathrm{i}}
\newcommand{\gr}{\gamma_\mathrm{r}}
% Green functions/tensors
\newcommand{\Gr}{\tens{G}}
\newcommand{\Grcavv}{\Gr_{\mathrm{c}}}
\newcommand{\Grhomv}{\Gr_{\mathrm{hom}}}
\newcommand{\Grbgv}{\Gr_{\mathrm{bg}}}
\newcommand{\Grbg}{G_{\mathrm{bg}}}
\newcommand{\Grhom}{G_{\mathrm{hom}}}
\newcommand{\Grcav}{G_{\mathrm{c}}}
\newcommand{\Grsv}{\tens{\mathbf{G}}_\mathrm{s}}
\newcommand{\Grsumv}{\tens{\mathcal{G}}}
\newcommand{\Grsumcv}{\tens{\mathcal{G}}_{\mathrm{c}}}
\newcommand{\Grsumhomv}{\tens{\mathcal{G}}_{\mathrm{hom}}}
%alpha
\newcommand{\al}{\tens{\alpha}}
\newcommand{\aleff}{\alpha_{\mathrm{eff}}}
\newcommand{\aldyn}{\alpha_\mathrm{dyn}}
\newcommand{\aldynv}{\al_\mathrm{dyn}}
%chi
\newcommand{\chic}{\chi_\mathrm{c}}
% electric/magnetic fields
\newcommand{\E}{\mathbf{E}}
\newcommand{\Es}{\E_\mathrm{s}}
\newcommand{\Eo}{\E_0}
\newcommand{\Ein}{\E_\mathrm{in}}
\newcommand{\Eintilde}{\tilde{\E}_\mathrm{in}}
\newcommand{\Ec}{E_\mathrm{c}}
\newcommand{\Etot}{E_\mathrm{tot}}
\newcommand{\Einyy}{E_\mathrm{in,yy}}
% mode profile
\newcommand{\ec}{\mathbf{e}_c}
% coordinates
\renewcommand{\r}{\mathbf{r}}
\newcommand{\ra}{\mathbf{r}_{\mathrm{a}}}
\newcommand{\rpar}{\r_{\parallel}}
\newcommand{\rparz}{\r_{\parallel,0}}
\newcommand{\ror}{\r_{\mathrm{or}}}
% constants
\newcommand{\ee}{\epsilon}
\newcommand{\eo}{\epsilon_0}
\newcommand{\mo}{\mu_0}
%Im and Re etc
\renewcommand{\Re}[1]{\ensuremath{\operatorname{Re}\left\{#1\right\}}}
\renewcommand{\Im}[1]{\ensuremath{\operatorname{Im}\left\{#1\right\}}}
% other physical quantities
\newcommand{\p}{\mathbf{p}}
\newcommand{\ptot}{p_{\mathrm{tot}}}
\newcommand{\Vm}{V_\mathrm{m}}
\newcommand{\Veff}{V_\mathrm{eff}}
\newcommand{\Celcius}{\ensuremath{^\circ}C\ }
\newcommand{\Deg}{\ensuremath{^\circ}\ }
\newcommand{\Fp}{F_{\mathrm{P}}}
% math entries
\newcommand{\ddt}[1]{\frac{\mathrm{d} #1}{\mathrm{d}t}}
\newcommand{\ddtsq}[1]{\frac{\mathrm{d}^2 #1}{\mathrm{d}t^2}}
\newcommand{\ii}{\mathrm{i}}
\newcommand{\jj}{\mathbf{j}}
\renewcommand{\d}[1]{\ensuremath{\operatorname{d}\!{#1}}}
%momenta
\newcommand{\kparv}{\mathbf{k}_{\parallel}}
\newcommand{\kparvin}{\mathbf{k}_{\parallel}^{\mathrm{in}}}
\newcommand{\kpar}{k_{\parallel}}
\newcommand{\kc}{k_{\mathrm{c}}}
\newcommand{\kcx}{k_{\mathrm{c},x}}
\newcommand{\kcy}{k_{\mathrm{c},y}}
\newcommand{\gv}{\mathbf{g}}
\newcommand{\kv}{\mathbf{k}}

%Fresnel
\newcommand{\rs}{r_{\mathrm{s}}}
\newcommand{\rp}{r_{\mathrm{p}}}
%Matrices
\newcommand{\MM}{\tens{M}}
\newcommand{\MO}{\tens{O}}

%%%%%%%%%%%

\onecolumngrid
\Large\textbf{Supplementary Material for ``Controlling nanoantenna polarizability through backaction via a single cavity mode"}
\normalsize

\section{Extinction in reflection}
\label{sec:extinction}
In our main paper we claim that the specular reflectance signal as we measure it in our experiment primarily reports on the imaginary part of the polarizability. 
We discuss and elucidate on this claim in this section. 

In Ref.~\cite{SIDeAbajo2007} De Abajo derives an expression (in cgs units) for the normal incidence specular reflection signal $r'$ that one expects from a particle array (with real space lattice area $A$) in free space. 
The observed reflection  depends on the self-consistent single particle polarizability $\alpha_\text{E}$, corrected by the lattice Green's function $G_{xx}(0)$, and reads
\begin{equation}
r'=\frac{2\pi\im k/A}{\alpha^{-1}_\text{E} - G_{xx}(0)}.
\end{equation}
Introducing the lattice dressed polarizability $\alpha^{-1}\equiv\alpha^{-1}_\text{E} - G_{xx}(0)$ and the background reflection $r_\text{glass}$ of our interface (see~\cite{SIChen2017} for an elaborate discussion on properties of plasmonic nanoantenna arrays on interfaces) we arrive at Eq.~(1) of our main text:
\begin{equation}
r' = r_\text{glass} + \frac{ 2\pi \im k \alpha}{A}.
\label{eq:rprime_Abajo}
\end{equation}
Generally, the Fresnel coefficient $r_\text{glass}$ is real valued.
Using this notion one can continue to write the specular reflectance $|r'|^2$ as
\begin{equation}
|r'|^2 = r_\text{glass}^2 - r_\text{glass} \frac{4\pi k}{A} \text{Im}[\alpha] + 
 \frac{4\pi^2 k^2}{A^2} |\alpha|^2,
\label{eq:rsq}
\end{equation}
which evidences that the imaginary part of the polarizability $\text{Im}[\alpha]$ leads to a reduction in specular reflectance, whereas the scattering term scaling with $|\alpha|^2$ results in an increased reflectance. 
Alternatively one can express Eq.~(\ref{eq:rsq}) as a function of the extinction cross section $\sigma_\text{ext}=4\pi k \text{Im}[\alpha]$ and scattering cross section $\sigma_\text{scat} = \frac{8}{3}\pi^2 k^4 |\alpha|^2$, which gives
\begin{equation}
|r'|^2 = r_\text{glass}^2 - \frac{r_\text{glass}}{A} \sigma_\text{ext} + \frac{3}{2} \frac{1}{A^2 k^2} \sigma_\text{scat}.
\label{eq:rsq_exsc}
\end{equation}
From this expression we learn that a reduction in reflectance, with respect to the background signal coming from the interface, can be associated with extinction.
For a plasmon particle or array such a reduction is expressed over a wide frequency range that is commensurate with its bandwidth (Fig.~\ref{fig:dipolemodel}c/d). 
In addition, Eq.~(\ref{eq:rsq_exsc}) shows that dilute lattices, such as the one we use, result in a more pure extinction measurement than dense lattices, for which the scattering term  contributes more strongly to the observed signal as a result of the larger proportionality factor $(A^2k^2)^{-1}$. 

\newpage
\section{Experimental details}
In this section we describe the fabrication of samples and the experimental setup. 
More details about the experimental arrangement can also be found in~\cite{SIRuesink2015}.

\begin{figure}
\center
\includegraphics[]{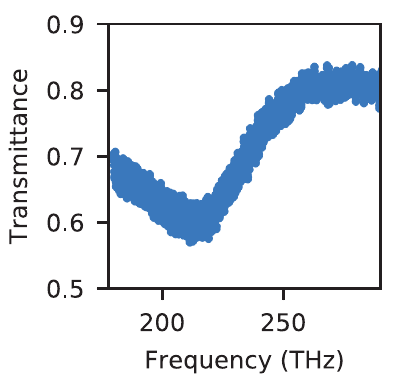}
\caption{Normal-incidence transmittance spectrum of the antenna array. 
}
\label{fig:norm_incidence}
\end{figure}

\subsection{Fabrication}
Gold nanoantennas are fabricated in an array~(\SI{150}{\micro\meter} by \SI{150}{\micro\meter}) on a $\SI{170}{\micro\meter}$ thick glass coverslide.
To start, a 130 nm layer of ZEP-520 resist is spin-coated on top of the coverslide.
The nanoantennas are patterned into the resist using electron beam lithography.
After development, thermal evaporation of gold and a lift-off step yield the desired antennas.
The antennas width and thickness were designed to be 120 and 40 nm and the length is approximately 400 nm. 
The pitch along the short axes of the antennas is 1500 nm, with a pitch along the long axes of 800 nm.
We characterize the spectral properties of the array (under normal incidence) using Fourier-Transform Infrared spectroscopy, obtaining a resonance frequency $\oa\approx\SI{208}{THz}$ and linewidth $\gamma$ of $\gamma\approx\SI{55}{THz}$ (Fig.~\ref{fig:norm_incidence} and \cite{SIRuesink2015}).

In our experiment we use a high Q silica microtoroid (diameter $\approx\SI{36}{\micro\meter}$) that is fabricated on the edge of a silicon sample.
For the fabrication protocol we largely followed methods that have been previously reported in for example~\cite{SIArmani2003,SIAnetsberger2010}.
In this work, spin-coating (ma-N 2410) and subsequent cleaving of the sample enabled targeted e-beam lithography of the disks on the edge of the sample.

\subsection{Response measurements: experimental setup and calibration}

\begin{figure}
\center
\includegraphics[]{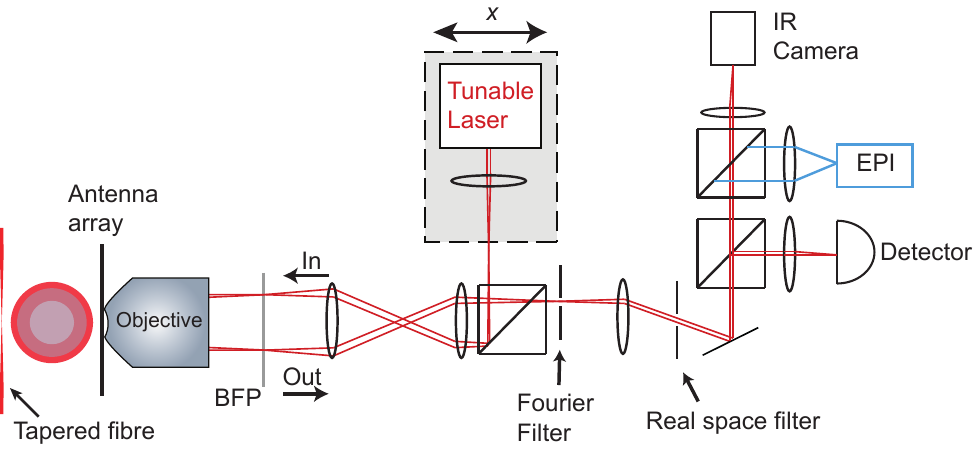}
\caption{Experimental setup that is used for response measurements on the array. 
See text for details. 
BFP: Back-Focal-Plane
EPI: EPI illumination using an infrared source.
}
\label{fig:setup}
\end{figure}

\begin{figure}
\center
\includegraphics[]{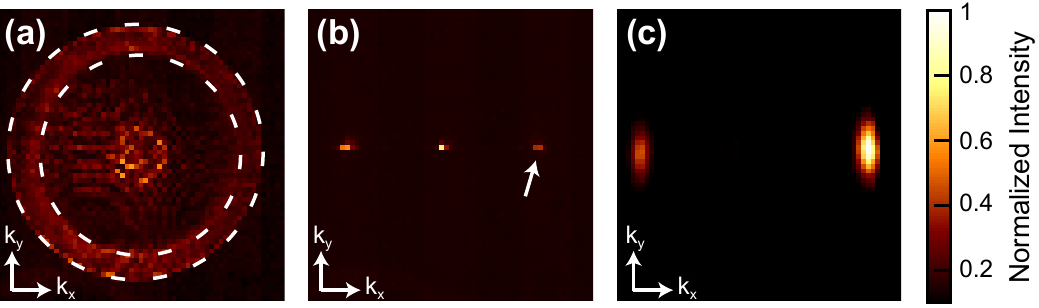}
\caption{
Fourier space images.
The color scale next to panel (c) is similar for panels (a)-(c).
(a) Epi-illumination on bare glass allows us to fit the NA=1 ($k_\parallel=k_0$) and maximum aperture circles.
The fitted NA=1 circle is used as a ruler for (b) and (c).
(b) Diffraction orders on the antenna array obtained via BFP laser illumination. 
(c) The cavity mode profile as obtained by directly exciting the cavity using an evanescently coupled tapered fiber.}
\label{fig:kspace_cal}
\end{figure}

A schematic of the optical setup used to probe the response of the array in the absence and presence of the microtoroid is depicted in Fig.~\ref{fig:setup}. 
The incident drive field is generated from a tunable laser source (New-Focus TLB-6728) and focused on the back-focal-plane (BFP) of a high-NA objective (Nikon, CFI Apo TIRF 100x) to illuminate our sample with a collimated laser beam.
Using a translation stage we control the position of the focus in the BFP, and thus the angle of incidence of the driving field. 
The reflection of the array back into the objective is Fourier filtered to ensure that only specular reflection is recorded on the detector.
In addition, the real space filter suppresses background singles of antennas that are not coupled to the cavity. 
The tapered fiber is used to directly excite the cavity mode (in a separate experiment) and obtain information on the cavity mode profile~(Fig.~\ref{fig:kspace_cal}c) and polarization of the cavity mode.
We also use the fiber to check that, with the cavity positioned in front of the glass substrate away from the antenna array, we do not directly excite the cavity mode with the incident drive field (and associated wavevectors) used in our experiment. 

\begin{figure}
\center
\includegraphics[]{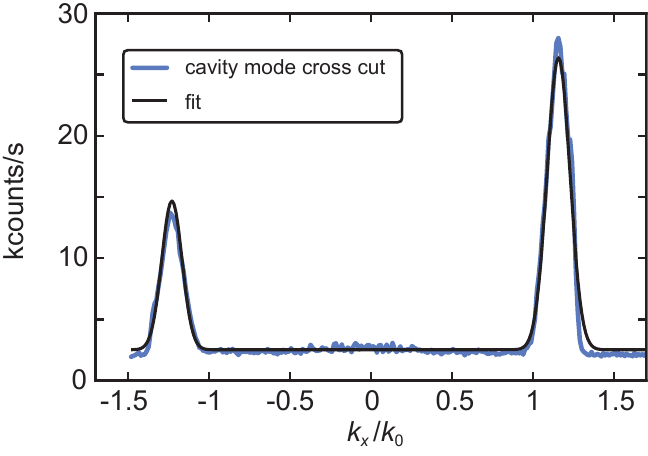}
\caption{Cross-cut of the cavity mode profile along the $k_x$ direction. The difference in intensity comes from the fact that we primarily drive one of the two modes using the evanescently coupled tapered fiber. 
We fit (black solid line)  both optical modes with a Gaussian lineshape.
}
\label{fig:kspace_cav}
\end{figure}

To generate the Fourier space images displayed in the main text we use three separately measured camera (Allied Vision Goldeye P008) images~(Fig.~\ref{fig:kspace_cal}). 
Using epi-illumination on bare glass we obtain information on the objective: The brighter red circular area on the edge of Fig.~\ref{fig:kspace_cal}a is associated with Total-Internal Reflection (high reflectance). 
Fitting the inner(outer) edge of this bright area with circular shapes (indicated by the dashed white lines) gives the position of the NA=1 ($k_\parallel=k_0$) and maximum aperture (NA$\approx$1.33) circles.
To obtain Fig.~\ref{fig:kspace_cal}b we excite the array using our tunable laser (focused in the BFP).
The resulting figure provides information on the diffraction orders on the antenna array and, together with the calibration (Fig.~\ref{fig:kspace_cal}a),  allows us to retrieve the wavevectors of specular and higher order diffraction. 
We do so by taking a cross-cut along the horizontal ($k_x$) direction.
The zero order reflection is indicated with an arrow and its position (obtained from a fit with  a Gaussian lineshape) provides information on the incident wavevector of the drive field. 
To generate the overlay of Fig.~2c (main text) we display the position of the diffraction orders using a grey saturated colour scale.
Fig.~\ref{fig:kspace_cal}c shows the radiation profile of the cavity mode.
To obtain this profile we position the cavity mode close to the bare glass substrate and directly excite the cavity using an evanescently coupled tapered fibre.
From a fit to a cross-cut along the horizontal direction (Fig.~\ref{fig:kspace_cav}) using a Gaussian lineshape, we obtain a cavity width in Fourier space of approximately $0.15 k_{||}/k_0$, with the cavity mode to which we couple via second order diffraction centered at $-1.23 k_x/k_0$.
Considering the incident free-space wavelength of 1540 nm and a pitch of 1500 nm, one would expect maximum coupling between the array and cavity (via the 2nd order diffraction) for an incident wavevector of $[-1.23+2\times(1540/1500)]=0.82k_x/k_0$, which matches relatively well with the experimentally observed incident wavevector $0.78k_x/k_0$ for which we observe our maximum in cooperativity.

% ---------  
\newpage

\section{Analytical dipole-dipole calculations}
\label{sec:alpha_latticesum}

In the main text we use a general coupled oscillator model to fit our experimental data.
In this section we provide additional support for our experimental results and the interpretation thereof using a full electrodynamic theory.
Our model treats each antenna in the array as a separate dipole and calculates the total response of the array using an analytical point-dipole model~(see for example Refs.~\cite{SIDeAbajo2007,SIKwadrin2016}). 
It is essential to understand that in such a coupled dipole model a dipole is driven not only by the driving field and it's own backscattered field, but also by the field scattered by the other dipoles.
For a lattice of $N$ identical scatterers, the dipole moment of particle $n$ reads
\begin{equation}
\p_{n}=\tens{\alpha}_0 \left[ \Ein(\r_{n}) + \sum_{m}^{N} \Gr(\r_{n},\r_{m},\oo) \p_{m}\right], \label{SI:Eq:CoupledDipoles1}
\end{equation}
with $\Ein(\r_{n})$ the driving field and $\tens{\alpha}_0$ the electrostatic particle polarizability.
The Green's function $\Gr=\Grbgv \!+ \! \Grcavv$ is the total Green's function, consisting, in our case, of the background and the cavity contributions, respectively. 
To gain more insight in our experimental results, we will solve \cref{SI:Eq:CoupledDipoles1} in two different ways.

First, in \cref{sec:finitelattice} we consider a finite array of scatterers placed in vacuum. 
To account for the presence of the ring resonator cavity, we derive an analytical expression for $\Grcavv$ that takes into account the specific geometry of a microtoroid ring resonator.
This approach has the benefit that it allows us to assign a position dependent Purcell factor that  each individual antenna in the array is subject to.  In other words,  the strength of this approach is that it can take into account important aspects of the cavity that include the cavity mode profile, Q and V,  as well as the fact that the toroid curvature means that only a finite set of particles are in its mode. 
While its strength is the description of the finite-sized cavity, its weakness is that it can only deal with a finite number of particles and can not account for the air-glass interface on which the particles are placed in the experiment. 

The second method we discuss (\cref{sec:infinitelattice}) is complementary as it assumes an infinite array of scatterers including all retarded electrodynamic interactions.   
However, because an infinite array requires Ewald summation in k-space, this method approximates the cavity as a translation invariant resonantly reflecting slab.
For an infinite periodic array, the polarizability is entirely summarized by the polarizability of a particle at the origin~\cite{SIDeAbajo2007}.
Importantly, it has recently been shown that the theory that typically describes such infinite arrays in vacuum~\cite{SIDeAbajo2007} can be extended to take into account a reflective surface on which the particles are placed~\cite{SIKwadrin2016}. 
As the resulting theory only requires Fresnel reflection and transmission coefficients, in fact one can even use stacked (resonant) planar layers as an interface~\cite{SIChen2017}. 
This is also the approach we take in this second model: we essentially lump the response of the interface and cavity into a single Fresnel coefficient, and calculate the response of the array using the resulting `engineered' metasurface.
As our calculation in this second scenario allows us to include  interfaces such as the glass-air interface that characterizes our sample surface, we are able to reproduce the Fano-lineshapes that are observed in our experiment (Fig.~4, main text). 
However, before we solve these two systems that consist of multiple dipoles, we first get acquainted with the most simple case: that of a single dipole coupled to an arbitrary single cavity mode.

\subsection{A single dipole coupled to a cavity: The relation between backaction and the Purcell factor}
\label{sec:Purcell}

For the limiting case of a single scatterer, the (radiative) decay rate enhancement ($\text{Im}[\tens{G}_\mathrm{c}]$, essentially the LDOS of the cavity~\cite{SINovotny2012}) offered by a single cavity mode directly relates to the cavity's Purcell factor~\cite{SIPurcell1946}. 
In this section we introduce the Purcell factor into a coupled oscillator formalism, taking the approach described in~\cite{SIDoeleman2016}, where both the scatterer and single cavity mode are treated as a harmonic oscillator. 

Consider an antenna with resonance frequency and linewidth $\oa$ and $\gamma$, respectively, described as a point dipole with dipole moment $\mathbf{p}=p\hat{\mathbf{p}}$, where we assume that it is only polarizable along $\hat{\mathbf{p}}$. It is driven by an incident field and coupled to a cavity with resonance frequency and linewidth $\ooc$ and $\kk$, respectively. 
The fields $\{E_\mathrm{in},E_\mathrm{c}\}$ are projections of the incident and cavity fields along $\hat{\mathbf{p}}$, respectively, at the location $\ra$ of the antenna. 
In the Fourier domain the resulting coupled equations read~\cite{SIDoeleman2016}
\begin{align}
(\omega_\mathrm{a}^2-\omega^2-\mathrm{i}\omega\gamma)p - \beta E_\mathrm{c} &= \beta E_\mathrm{in},
\label{eq:EOM_ant_F}  \\
-\frac{\omega^2}{\epsilon\epsilon_0 V_\mathrm{eff}}p+(\omega_\mathrm{c}^2-\omega^2-\im\omega\kappa)E_\mathrm{c} &= 0, 
\label{eq:EOM_cav_F}
\end{align}
where $V_\mathrm{eff}$ is the effective mode volume of the cavity as it is felt by the antenna at position $\rr_0$~\cite{SIDoeleman2016}, $\epsilon=\epsilon(\ra)$ and $\beta$ is the antenna's  oscillator strength.
From \cref{eq:EOM_ant_F} we recognize the antenna's response  $\alpha_\mathrm{dyn}$ as  $\alpha_\mathrm{dyn}={\hat{\p} \cdot\! \aldynv \!\cdot\hat{\p}} =\beta/(\oa^2-\oo^2-\im\omega\gamma)$. 
%and similarly ${G_\mathrm{c} = \hat{\mathbf{p}}\cdot\ \!\! \tens{G}_\mathrm{c} \! \cdot \hat{\mathbf{p}}}$. 
Note that here the response of the antenna $\aldyn$ is the single particle polarizability that is corrected for the Greens function of the embedding environment $\Grbgv$ \textit{excluding} the single cavity mode. 

Combining \cref{eq:EOM_ant_F,eq:EOM_cav_F} yields an expression for the effective polarizability $\aleff$, defined through $p= \aleff E_\mathrm{in}$, which reads
\begin{equation}
\aleff = \left[\alpha_\mathrm{dyn}^{-1} - \chi_\mathrm{c} \right]^{-1}.
\label{eq:alpha_eff1}
\end{equation}
Here we have defined the cavity's response function $\chic$ via 
\begin{equation}
\chic  \equiv \frac{E_\mathrm{c}}{p} = \frac{1}{\epsilon\epsilon_0 V_\mathrm{eff}}\frac{\oo^2}{\oc^2-\oo^2-\im\oo\kappa}.
\label{eq:cav_field}
\end{equation}

On the other hand, for a single dipole with one polarization axis we can also rewrite \cref{SI:Eq:CoupledDipoles1} as $p= \aleff E_\mathrm{in}$, where 
\begin{equation}
\aleff = \left[\aldyn^{-1} - \Grcav(\r_0, \r_0,\oo) \right]^{-1}.
\label{eq:alpha_eff2}
\end{equation}
Here, $\Grcav=\hat{\p} \cdot \Grcavv \cdot \hat{\p}$ and we have used that $\aldynv=\left[\tens{\alpha}_0^{-1} - \Grbgv \right]^{-1}$. Comparing \cref{eq:alpha_eff1,eq:alpha_eff2}, we recognize that $\chi_\mathrm{c}=\Grcav (\r_0,\r_0,\oo)$. 
We can now make a connection to the Purcell factor $F$, given by~\cite{SIPurcell1946}
\begin{equation}
F=\frac{6\pi}{k^3}\frac{Q}{V_\mathrm{eff}},
\label{eq:alpha_Purcell}
\end{equation}
with $k=n\oo/c$, $n$ the refractive index of the medium embedding the scatterer and the quality factor $Q=\oc/\kk$. It is well known that $F$ describes the emission rate enhancement that a dipolar emitter experiences at resonance with a cavity. If we consider the response of the cavity at $\oo=\oc$, we find
\begin{equation}
\chic \overset{\oo=\oc}{=} \frac{\im}{\epsilon_0\epsilon}\frac{Q}{V_\mathrm{eff}} = \frac{\im k^3}{6\pi\epsilon_0\epsilon} F.
\label{eq:chiatres}
\end{equation}
Using this relation and fact that $\gamma=\gi+\gr$, with $\gi$ the Ohmic losses and $\gr$ the radiative losses given as $\gr=\beta \oo^2 n^3/(6 \pi c^3 \epsilon_0\epsilon)$ in a homogeneous medium~\cite{SIDoeleman2016}, we can rewrite \cref{eq:alpha_eff1} for $\oo=\oc$ as
\begin{equation}
\aleff \overset{\oo=\oc}{=} \frac{\beta}{\oa^2-\ooc^2 - \im\ooc\gamma_0 -
\mathrm{i}\ooc \gamma_\mathrm{r}[1 +  F ] }
\label{eq:alpha_PurcellAnt}
\end{equation}
This important result shows that, like a dipolar emitter, a single scatterer coupled to a single mode cavity experiences an enhanced \textit{radiative} loss rate that is directly given by the Purcell factor $F$ (for $\oo=\oc$). When antenna and cavity are close in resonance frequency, the effect of this enhanced loss rate is that the antenna polarizability $\aleff$ is strongly suppressed around the cavity resonance \cite{SIFrimmer2012}. 
The strength of this suppression is thus a measure for Purcell factor. 
This result and the connection to cooperativity and Albedo is further discussed in \cref{sec:Albedo}.

\subsection{A finite dipole lattice with analytical cavity Green's function}
\label{sec:finitelattice}

In our experiment we did not probe the response of a single scatterer, but instead measured on an array of dipoles. 
Here we use a brute-force coupled dipole model to show that the response of an array qualitatively matches the response of a single scatterer when coupled to a single cavity mode. 
The spectral lineshapes that we calculate in both scenarios are similar, although lattice effects can lead to a significantly stronger response for some particles in the array, compared to that of the single scatterer case. 
We first introduce the coupled dipole model, before deriving the Greens function of a single whispering gallery mode (WGM) and finally showing the results that we obtain using our model. 

\subsubsection{Retrieval of $\aleff$ in a finite lattice}

For $N$ dipoles, \cref{SI:Eq:CoupledDipoles1} leads to a set of $3N$ coupled equations of motion. To simplify the math, we take the particles to be only polarizable along the y-axis, reducing \cref{SI:Eq:CoupledDipoles1} from $3N $ to $N$ equations. Reshuffling the terms, we can now write an equation of the form
\begin{equation}
\tens{M}^{-1}\mathbf{\tilde{P}} = \Eintilde
\end{equation}
with
\begin{equation}
\tens{M}^{-1}=
	\begin{pmatrix}
    \alpha_{yy}^{-1} - G_{yy}(r_{0},r_{0}) & - G_{yy}(r_{0},r_{1}) & ... & - G_{yy}(r_{0},r_{N}) \\
     - G_{yy}(r_{1},r_{0}) & \alpha_{yy}^{-1} - G_{yy}(r_{1},r_{1}) & ... & \vdots \\
     \vdots &  & \ddots & \\  
     - G_{yy}(r_{N},r_{0}) & ... & & \alpha_{yy}^{-1} - G_{yy}(r_{N},r_{N})\\
	\end{pmatrix} \label{Eq:InvM}
\end{equation}
and where $\mathbf{\tilde{P}}$ and $\Eintilde$ are column vectors of length $N$ containing the dipole moments of all particles and the driving fields at their positions, respectively. We can solve this system of equations by setting up $\tens{M}^{-1}$ and numerically inverting it. 
One then multiplies it with the driving fields $\Eintilde$ to get $\mathbf{\tilde{P}}$. Dividing $\mathbf{\tilde{P}}$ element-wise by $\Eintilde$, one obtains the effective polarizability $\alpha_{\mathrm{eff}}$ of each particle, defined as usual through $p_n=\aleff E_{\mathrm{in}}(\r_n)$.

\subsubsection{The cavity Green function}

In \cref{sec:Purcell} we related the Purcell factor and the cavity response $\chic$ to an arbitrary $\Grcav (\ra,\ra,\oo)$, the cavity Green function at the location of a scatterer. 
However to couple multiple dipoles via the cavity, we can not simply deal with a generic expression for $\Grcav (\ra,\ra,\oo)$, but instead require an explicit expression that describes or approximates  the full cavity Green function.
Such an expression necessarily includes the cavity mode profile.

The field of a single cavity mode can be described as \cite{SIDoeleman2016}
\begin{equation}
\E(\r, \omega) = a(\omega) \ec(\r), \label{SI:Eq:Eigenfield}
\end{equation}
where $a(\omega)$ is the frequency-dependent amplitude and $\ec(\r)$ is the normalized field profile. 
We can set up an equation of motion for $a(\omega)$ including a drive dipole $\p'$ at position $\r'$, similar to \cref{eq:EOM_cav_F} \cite{SIDoeleman2016}. 
Solving this equation and taking the small linewidth approximation ($\kk \ll \ooc$), we get 
\begin{equation}
a(\omega) = \frac{i}{4} [\ec^*(\r') \cdot \p'] \frac{\ooc}{-i\Delta + \kappa/2}, \label{SI:Eq:a1}
\end{equation}
with $\Delta = \omega-\ooc$. 
Here $\ec^*$ denotes the conjugate transpose of $\ec$. 
The cavity Green's function is defined through the cavity fields generated at position $\r$ by a dipole at position $\r'$ as
\begin{equation}
\E(\r)=\Grcav(\r,\r',\omega) \p'.
\end{equation}
Plugging in \cref{SI:Eq:Eigenfield,SI:Eq:a1} for $\E(\r)$ and reshuffling terms, we get
\begin{equation}
\Grcav(\r',\r,\omega) = L(\omega) \, \ec(\r) \ec^*(\r') \,  , \label{SI:Eq:Grcav1}
\end{equation}
with $L(\omega) = \frac{i}{4} \frac{\ooc}{-i\Delta + \kappa/2} $ the Lorentzian lineshape function. Let us note that the modes are normalized such that 
\begin{equation}
\frac{1}{2} \iiint \eo \ee(\r) |\ec(\r)|^2 dV =1
\end{equation}
and that the effective mode volume experienced by a dipole $\p'$, positioned at $\r'$, as it is used e.g. in \cref{eq:EOM_cav_F}, is defined as
\begin{align}
\Veff(\r') &=\frac{ \iiint \eo \ee(\r) |\ec(\r)|^2 dV}{ \eo \ee(\r') |\hat{p}' \cdot \ec(\r')|^2 } \\
&=\frac{2}{\eo \ee(\r') |\hat{p}' \cdot \ec(\r')|^2 } \label{SI:Eq:Vmode1}
\end{align}

\begin{figure}[tb]
\begin{center}
\includegraphics[width=\figwidth]{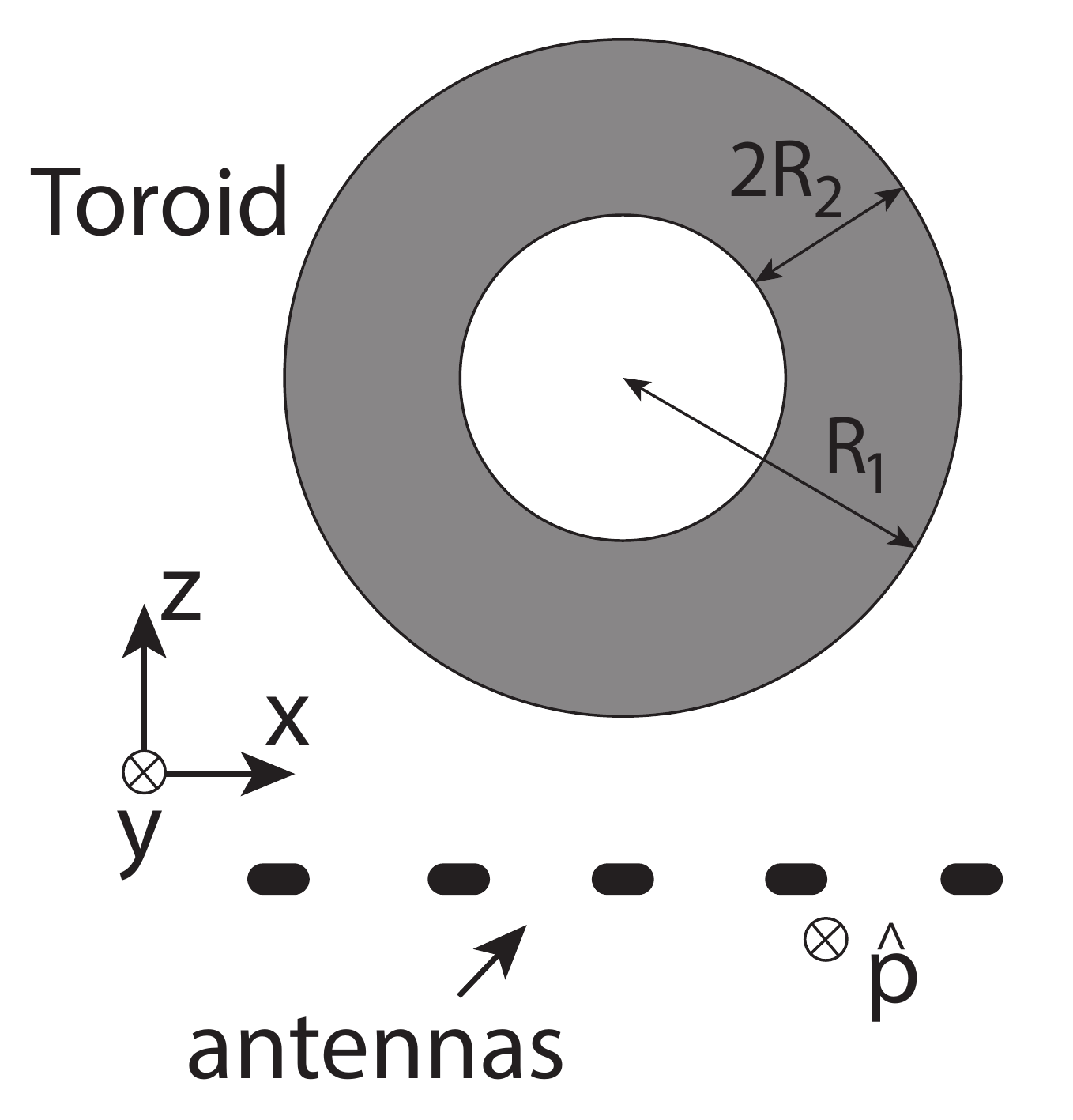} 
\caption{A 2D array of antennas near a cavity. The antennas form a lattice in the x-y plane. The antennas have their dipole moments oriented along the y-axis. The cavity is a microtoroid with major and minor radii $R_1$ and $R_2$, respectively.} \label{Fig:Sketch} 
\end{center}
\end{figure} 

From \cref{SI:Eq:Grcav1} it is clear that we need to find an expression for the spatial mode profile of the cavity. It was shown that the mode profile of a fundamental TE WGM in a microtoroid with major radius $R_1$ and minor radius $R_2$, outside the glass of the cavity, takes on the shape \cite{SIAnetsberger2010}
\begin{equation}
\ec(\r)=\ec(y=0,r=R_1) e^{-y^2/2r_y^2} e^{-\kk_r (r-R_1)} e^{\pm i l \phi}, \label{Eq:ModeProfile1}
\end{equation}
where $\ec(y=0,r=R_1)$ is the field somewhere on the edge of the cavity, in the equatorial plane, $r_y$ is a Gaussian width along the $y$-direction (which depends on $R_1$, $R_2$ and $l$), $\kk_r\approx \frac{2 \pi \sqrt{\epsilon_c-\epsilon}}{\lambda}$ (with $\epsilon_c$ and $\epsilon$ the permittivities of the cavity and the surrounding, respectively) is the radial decay length and $l$ is the azimuthal mode number. $r$ and $\phi$ are cylindrical coordinates, where we have taken the origin to lie in the toroid center. See \cref{Fig:Sketch} for a sketch of the geometry. A plus or a minus sign in the azimuthal dependence determines whether it is the counterclockwise (CCW) or clockwise (CW) mode in the toroid. Note that the total cavity Green's function is the sum of the CW and CCW mode contributions. Since we only want to know the field in the plane of the antennas, close to where the toroid approaches the lattice, we can make a Taylor expansion around $x=0$ in the Gaussian term. Doing the same in the last term describing the azimuthal dependence, we get for the field in the plane of the antennas
\begin{equation}
\ec(\r)=\ec(\{x,y\}=0,z=R_1) e^{-x^2/2r_x^2} e^{-y^2/2r_y^2} e^{-\kk_r (|z|-R_1)} e^{\pm i k_c x}, \label{Eq:ModeProfile2}
\end{equation}
where $r_x=\sqrt{z/4 \kk_r}$ and $k_c = -l/z$ is the effective wavevector of the cavity mode in the antenna plane.

The cavity Green's function in the $z=z_0$ plane of the antennas can now be described as
\begin{equation}
\Grcav(\r',\r,\oo) = L(\oo) \MO e^{-(x^2+x'^2)/2r_x^2} e^{-(y^2+y'^2)/2r_y^2} e^{\pm i k_c (x-x')} . \label{SI:Eq:Grcav2}
\end{equation}
where $\r'$ and $\r$ are the source and detection positions, respectively, and
\begin{align}
\MO &= \ec(\{y,x\}=0, z=R_1) \ec^*(\{y,x\}=0, z=R_1) e^{-2 \kk_r (|z_0|-R_1)}  \\
&= \ec(\r_0) \ec^*(\r_0)
\end{align}
is a matrix with the fields at the origin $\r_0$ of the lattice.

\subsubsection{Results}

For simplicity, we restrict ourselves to a lattice of scatterers in vacuum.
The Greens function of the background $\Grbgv(\r_{n},\r_{m},\oo)$ is then a well-known expression~\cite{SINovotny2012}. 
We ignore its real part for $\r_m = \r_n$, taking the divergent electrostatic contribution to be included in the polarizability.
The cavity Greens function is the sum of the CW and CCW contributions
described in \cref{SI:Eq:Grcav2}. 
Since we assume the scatterers are only polarizable along the y-axis, we only require the yy-component of $\MO$. 
Using \cref{SI:Eq:Vmode1}, we can relate this to the effective mode volume $\Veff(\r_0)$ felt by a y-oriented dipole at the lattice origin, as $\MO_{yy}= 2/(\Veff(\r_0) \eo\ee(\r_0))$.

We assume particles with a Lorentzian polarizability $\alpha_0$ along the y-axis only, with resonance frequency $\oa=2\pi c/ \lambda_{\mathrm{a}}$, $\lambda_{\mathrm{a}}=$\SI{1.5}{\micro\meter}, and an ohmic damping rate of $\oa/10$, matching literature values of gold \cite{SIPenninkhof2008}. The dipoles are positioned in a square lattice with pitches $d_x=$\SI{1.5}{\micro\meter} and $d_y=$\SI{0.8}{\micro\meter}, containing 465 particles, i.e. 31 (15) particles in the x-(y-)direction. Our cavity is made of glass ($n=1.5$) and surrounded by air, has a major radius of \SI{18}{\micro\meter}, $l$=100, $\ooc=2\pi c / \lambda_c$ with $\lambda_c=\SI{1.5}{\micro\meter}$, $Q=3 \cdot 10^6$ and is located at \SI{2}{\micro\meter} distance from the lattice. This leads to a cavity in-plane wavevector of $k_c= 1.19 \, k_0$ (in good correspondence with the experimentally observed $1.23 \, k_0$) and a Gaussian width $r_x\approx \SI{1.46}{\micro\meter}$, and we take $r_y=r_x/2.6$, corresponding to the measured cavity mode profile. We choose $\Veff(\r_0)=5 \cdot 10^4 \lambda_c^3$ for the CW and CCW modes, meaning that the cavity Purcell factor at the lattice origin is 9.1. We excite the lattice with a plane wave at an angle with the normal, along the x-axis, i.e. $\kparv=\kpar \hat{\mathbf{x}}$.

\begin{figure}[htb]
\begin{center}
\includegraphics[width=\linewidth]{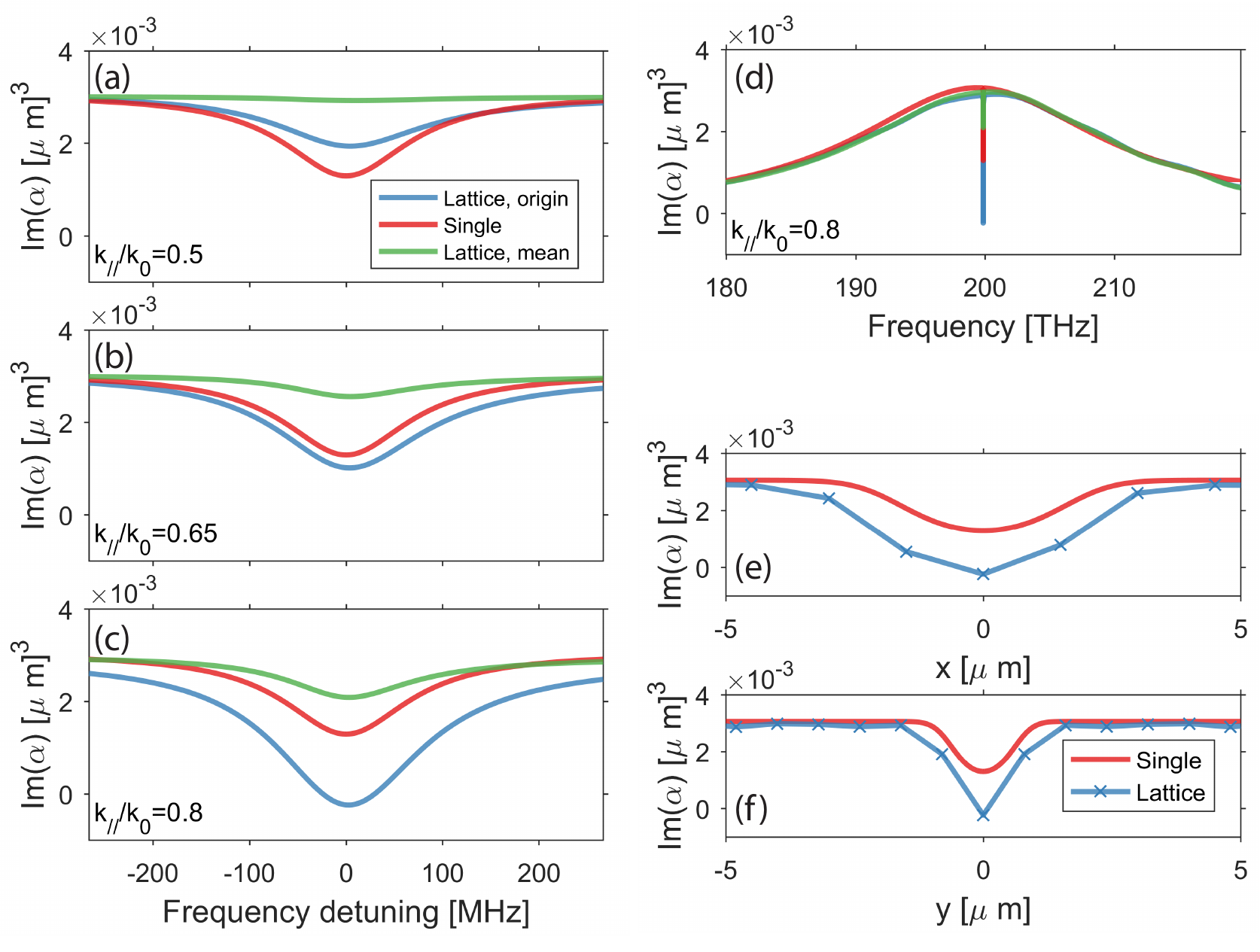} 
\caption{Effective polarizability $\alpha$ in a finite lattice of dipolar scatterers coupled to a WGM cavity. 
(a-c) Narrowband spectra of $\Im{\alpha}$, for three different incident parallel wavevectors $\kpar$. We show $\Im{\alpha}$ for a single dipole (located at the lattice origin $\r_0$) without a lattice (red) and for dipoles in a lattice, where we show the dipole at the lattice origin (blue) and the mean value of the dipoles within a spatial region of \SI{4.5}{\micro\meter} diameter around the origin (green).
(d) Broadband spectrum of $\Im{\alpha}$ for $\kpar/k_0=0.8$, where we have optimal phase matching to the cavity mode via the (-2,0) diffraction order. Color coding is the same as in (a-c).
(e-f) Spatial profiles of $\Im{\alpha}$ in the x- and the y-direction, centered at the lattice origin. We show $\Im{\alpha}$ for a single scatterer that is moved in the plane of the lattice (red) and for the particles in the lattice (blue). We chose $\oo=\ooc$ and $\kpar/k_0=0.8$.
Note that in these figures we use CGS units for $\alpha$, i.e. $\alpha_{\textrm{CGS}}=4\pi\epsilon_0 \epsilon \, \alpha_{\textrm{SI}}$, where $\alpha_{\textrm{SI}}$ is the polarizability in SI units, as used in our equations.} \label{Fig:SI-FiniteLattice} 
\end{center}
\end{figure} 

\Cref{Fig:SI-FiniteLattice} shows the effective polarizability of particles in this lattice. 
From \cref{Fig:SI-FiniteLattice} (a-c) we can see firstly that, while the polarizability of a single particle coupled to a cavity does not depend on angle of incidence, that of particles in a lattice does. 
We see exactly the type of phase-matching condition that was also observed in our experiment (Fig.~4 in the main text), where the effect of the cavity on polarizability is strongest when we are phase-matched to the cavity mode via the (-2,0) diffraction order. 
Here this occurs at $\kpar/k_0=0.8$. 
Moreover, we see that if phase-matching is achieved, the backaction effect introduced by the cavity can be stronger in a lattice than for a single particle: in a lattice, the particle at the origin has a more strongly modified polarizability than the single particle. 
This is because of the constructive interference of all particles radiating into the cavity, leading to an enhanced backaction field for particles near the origin. 
\Cref{Fig:SI-FiniteLattice} (d) displays the broadband polarizability of the dipoles, showing that both the single particle and the lattice follow the same lorentzian lineshape outside the cavity spectral window.
In \cref{Fig:SI-FiniteLattice} (e-f) we see that the particles close to the origin are more strongly affected by the cavity, \textit{i.e.,} that the effect diminishes for antennas at larger distance from the origin, roughly following the 2D Gaussian profile of the cavity mode.  
To compare the analytical results with our experiment, we also show in \cref{Fig:SI-FiniteLattice} (a-d) the mean polarizability of particles within a spatial region of \SI{4.5}{\micro\meter} diameter around the origin, corresponding to the size of the real-space filter used in our experiments. 
To first order, the field scattered by the dipoles within the filter area is proportional to their average polarizability. 
We see that this shows the same line shape, but the averaging decreases the effect of the cavity. 

To conclude, we have seen that the polarizability in a finite lattice of dipoles is qualitatively similar to the polarizability of a single dipole. 
We can thus use the ratio between the averaged lattice response and that of the single particle to predict the response we would have obtained in our measurement if we would have measured on a single particle, instead of on an array of dipoles. 
We do this prediction in \cref{sec:Albedo}, where we also discuss the relation between Albedo, cooperativity and Purcell factor. 
Finally we point towards an important difference between the single particle and lattice response: in a lattice, the backaction induced by the cavity mode can be tuned in strength by changing the angle of incidence, which as expected is not possible for a single scatterer. 

\subsection{Infinite lattice}
\label{sec:infinitelattice}

In this section we will discuss an infinite array of scatterers in front of an interface. 
With this model we aim to justify a specific claim made in the main text, namely that the Fano lineshapes that we observe in our experiment result from non-trivial background signals originating from the interface. 
Before we move to the details of our model, we first recap the foundations of the theory that describes infinite arrays of scatterers (see for example \cite{SIDeAbajo2007,SIKwadrin2016}).

The dipole moment $\mathbf{p}$ for a particle placed at the origin of an infinite array at a distance $d$ from an interface, that is excited by a plane wave with parallel wave vector $\mathbf{k}_\parallel$, is written as
\begin{equation}
\mathbf{p}=\left[\tens{\alpha}_\mathrm{int}^{-1} - \sum_{n\neq 0} 
\tens{G}_\mathrm{b}\!((\mathbf{R}_n,d),(0,d)) \, e^{\im\mathbf{k}_\parallel\cdot \mathbf{R}_n}  \right]^{-1} \mathbf{E}_\mathrm{in}'.
\label{eq:alpha_response_lattice}
\end{equation}
Here the sum over $\tens{G}_\mathrm{b}$ indicates the field at the origin generated by all particles, excluding the field generated by the dipole placed at this position.
Furthermore,  $\mathbf{R}_n$ is the real space lattice vector for each dipole $n$ (indicating its position) and $\mathbf{E}_\mathrm{in}'$ the incident field including its reflection at the interface. 
Importantly, note that the single particle polarizability $\tens{\alpha}_\mathrm{int}$ in this formula is already corrected for its homogeneous (embedding) environment \textit{and} the interface~\cite{SINovotny2012}, but not its neighbouring dipoles.  
The lattice summed Green function $\mathcal{G}$ over all point-dipoles except the origin ($n\neq 0$) is then defined as
\begin{equation}
\mathcal{G}^{\neq}\equiv
\sum_{n\neq 0} \tens{G}_\mathrm{b}\!((\mathbf{R}_n,d),(0,d)) \,  e^{\im\mathbf{k}_\parallel\cdot \mathbf{R}_n} ,
\label{eq:alpha_Glatt}
\end{equation}
which should be solved to find the response of the lattice. 
We will not discuss in detail on how to do this, but instead point to the relevant literature (see \textit{e.g.}~\cite{SILinton2010,SIKwadrin2016} for details).
The formalism as it is described by \cref{eq:alpha_response_lattice,eq:alpha_Glatt} is well established for 2D lattices in homogeneous space, using Ewald summation for exponential convergence of the lattice sums in the case that $\tens{G}_\mathrm{b}$ is $\tens{G}_\mathrm{hom}$ (with $\tens{G}_\mathrm{hom}$ the Green function for homogeneous space)~\cite{SIDeAbajo2007}. 
Recently, Kwadrin \textit{et al.}~\cite{SIKwadrin2016} showed how to generalize this approach for the case of lattices placed in front of a single reflective interface.
In this case, one separates $\tens{G}_\mathrm{b}$ as the sum of $\tens{G}_\mathrm{hom}$ and a reflected Green function $\tens{G}_\mathrm{refl}$, where $\tens{G}_\mathrm{refl}$ is written in the angular spectrum representation, taking the wave vector dependent Fresnel coefficient as an input.
After solving for $\mathcal{G}^{\neq}$, the subsequently obtained lattice- and interface-corrected polarizability $\alpha_\mathrm{lat}$ can used to calculate far-field observables such as, for example, reflection and transmission properties~\cite{SINovotny2012}. 

\subsubsection{Simple model for cavity interaction}

\begin{figure}
\center
\includegraphics{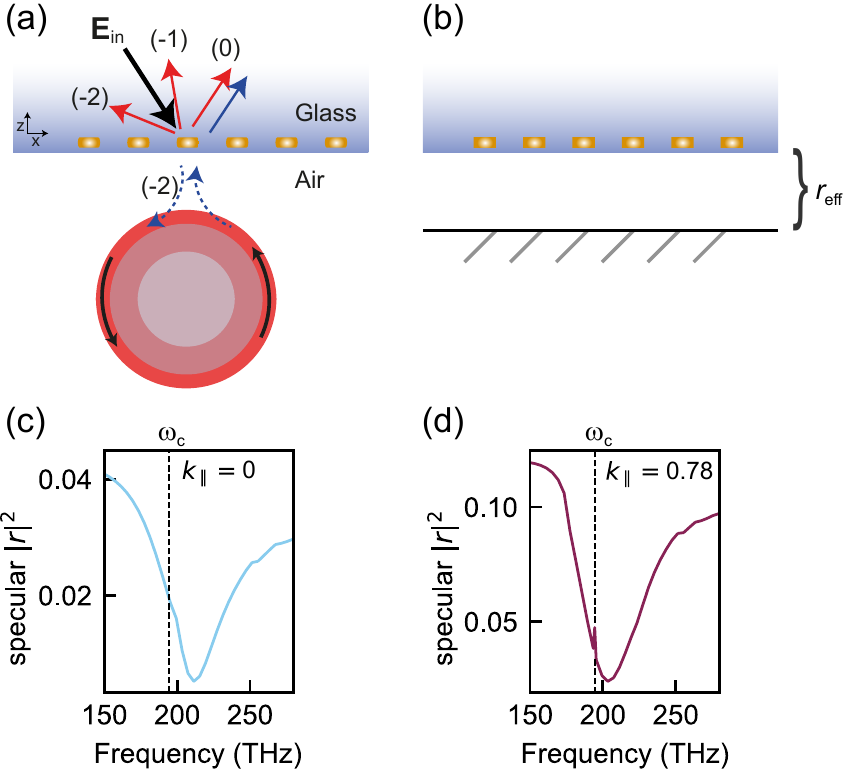}
\caption{\textbf{Lattice-sum calculations.}
(a) Cartoon explaining the position of the antennas in the lattice-sum calculation. 
In contrast to the experiment, the antennas are placed inside the glass (which is given a refractive index of 1.5) environment.
(b) In our calculation, we capture the combined effect of the interface and cavity in a single Fresnel coefficient $r_\mathrm{eff}$. 
(c) The calculated specular reflectance of the array (for normal incidence) shows a clear broadband dip associated with the plasmon resonance. 
The dashed line indicates the resonance frequency associated with the cavity mode.
(d) For $k_\parallel/k_0=0.78$, we observe the same signature. 
Note that due to the larger angle of incidence, the background signal coming from the glass-air interface is higher than in (c).
Importantly, the sharp feature that is visible at $\omega_\mathrm{c}$ is directly related to the presence (backaction) of the cavity. 
A calculation exploring this effect over a more narrow bandwidth is given in Fig.~\ref{fig:dipolefull}.
}
\label{fig:dipolemodel}
\end{figure}

Due to the curvature of the microtoroid cavity, lattice sum theory that relies on wave vector conservation arguments, as is the case when considering infinite lattices, is not strictly applicable to our experiment. 
To nonetheless approximate the experiment, we propose to mimic the cavity response by a resonant planar structure. 
This is a feasible approach, because the extension of lattice sum theory to lattices near mirrors (as reported by~\cite{SIKwadrin2016}) is not restricted to a single reflective interface. 
Instead, one can also consider an array of scatterers positioned in a half space in front of an arbitrary multi-layer stack~\cite{SIChen2017}.
This approach simply relies on replacing the Fresnel coefficient of the single interface with the multi-layer reflection coefficient $r_\mathrm{eff}$.

We refer to \cref{fig:dipolemodel}a/b for the proposed multilayer description.
The antenna array is positioned at \SI{50}{nm} from the interface, and the reflection coefficient of the stack is given by  
\begin{equation}
r_\mathrm{eff} = r_\mathrm{cav} + r_\mathrm{glass}.
\end{equation}
Due to the resonant nature of the single cavity mode (in frequency and wave vector), $r_\mathrm{eff}$ is equivalent to $r_\mathrm{glass}$, except for very specific frequencies and wave vectors at which it is possible to excite the cavity. 
For our calculations, we approximate $r_\mathrm{cav}$ as
\begin{equation}
r_\mathrm{cav} \equiv \frac{-\kappa_\mathrm{ex}}{(\omega-\omega_\mathrm{c}) + \im\kappa/2},
\label{eq:alpha_rcav}
\end{equation}
with $\kappa_\mathrm{ex}$ defined as 
\begin{equation}
\kappa_\mathrm{ex} \equiv \frac{\kappa}{2} \times e^{(-|k_\parallel-k_{\parallel,\mathrm{c}}|^2)/(2\sigma^2)} .
\label{eq:alpha_kex}
\end{equation}
Here $\kappa_\mathrm{ex}$ constitutes a Gaussian lineshape centered at  $k_{\parallel,\mathrm{c}}$ (the wavevector of the cavity mode) with a width given by $\sigma$. 
As a result of these definitions, $r_\mathrm{cav}$ is only nonzero over a small (Lorentzian) frequency bandwidth and for particular wave vectors, in close analogy the experimental situation and to the analytically derived  Green's function for the microtoroid that is displayed in a compact way in  \cref{SI:Eq:Grcav1}.

We note that the pre-factor $\kappa/2$ in \cref{eq:alpha_kex} is chosen such that \cref{eq:alpha_rcav} yields unity reflection for perfect phase-matching and $\omega=\omega_\mathrm{c}$, maximizing the effect of our resonant structure.
A drawback of our model is that we can not easily determine the `real' pre-factor that we should use.
In reality, the pre-factor should relate to the cavity-array distance, and determine the strength of the backaction. 
This is easily realized when taking the limit of infinite cavity-array separation, then $r_\mathrm{eff}$ should be entirely given by $r_\mathrm{glass}$ such that $r_\mathrm{cav}$ is always 0.
To end this section, we mention two other approximations. 
First of all, we note the resonant planar structure has an equal interaction with all antennas in the array. 
Although this is a requirement for the proper implementation of our calculations, it is in contrast to our experiment, where the curvature of the microtoroid ensures that the single cavity mode only interacts with a select number of antennas. 
A second difference between experiment and calculation is the positioning of the antenna array inside the glass environment, which is necessary to approximate the interface+cavity as a simple reflective multilayer.
This positioning will slightly influence the total field at the position of the array. 
However, taking into account that s-polarized fields are continuous across the boundary, and that we positioned our antennas at a distance of $\lambda/20$ from the substrate, we estimate the resulting difference originating from this change in position to be relatively small. 

\begin{figure}
\center
\includegraphics{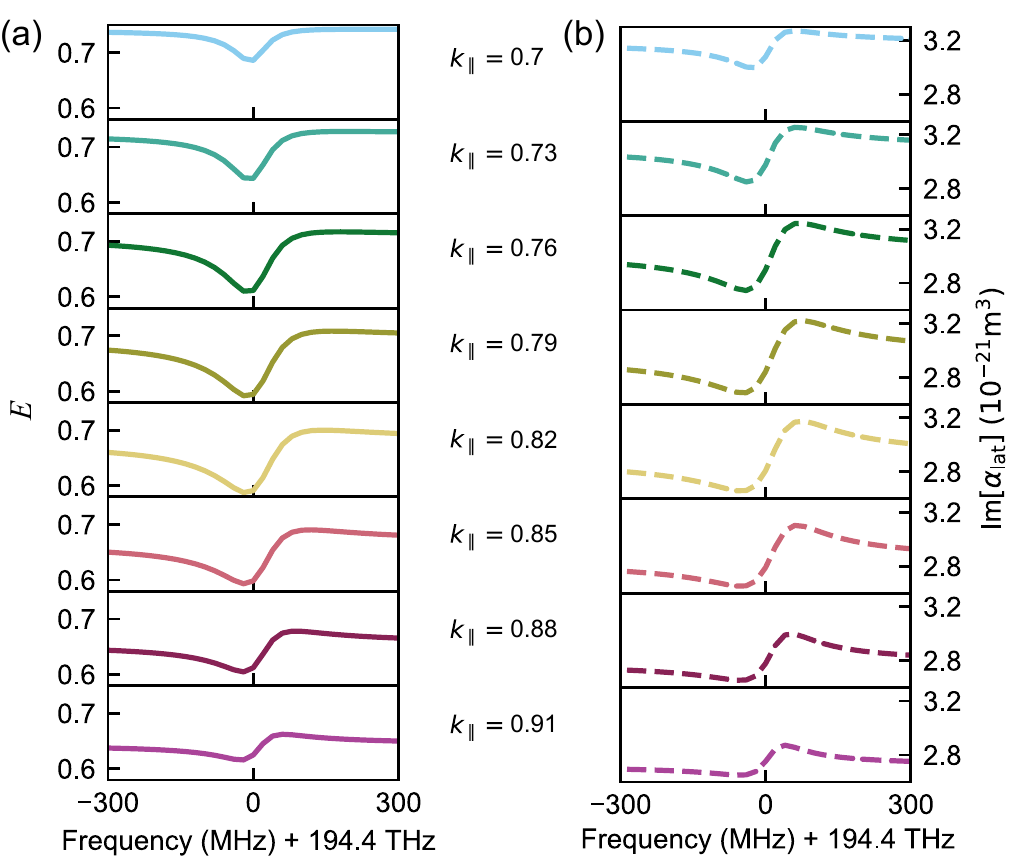}
\caption{
For both (a) and (b) the sub panels are associated with incoming wave vectors (in units of $k_0$) given by the values in between (a) and (b).
(a) Similar to our experiment, the calculation predicts a narrow bandwidth dip in reflectance resulting from backaction, that is maximized for a wave vector for which the second diffraction order matches the wave vector of the cavity. 
(b) The modification of $\text{Im}[\alpha_\mathrm{lat}]$ also displays this strong dependence on incoming wave vector. 
Interestingly, its shape is more reminiscent of a Fano lineshape. 
We attribute this to the fact that we do not perform a pure extinction measurement, solely probing $\text{Im}[\alpha_\mathrm{lat}]$, but are instead also sensitive to a term scaling with $|\alpha_\mathrm{lat}|^2$.
}
\label{fig:dipolefull}
\end{figure}

\subsubsection{Calculation results}
Using our model, we plot the specular reflectance spectra for two different illumination conditions in \cref{fig:dipolemodel}c (normal incidence) and \cref{fig:dipolemodel}d ($k_\parallel/k_0=0.78$).
In both scenarios we observe a broadband dip together with  a glass related background reflection signal.
This means that the reduction in reflection can be attributed to the plasmon resonance. 
Our calculations thus validate our claim that a dip in reflectance is a measure for extinction, and that an increase in reflectance signals a \textit{reduction} in extinction.
Moreover, it is noteworthy that these analytical calculations show very good agreement with the specular reflectance spectrum as predicted in the main text (Fig.~3a).

In addition to this broadband dip, we observe a small peak in the reflectance spectrum that is associated with the presence of the cavity in \cref{fig:dipolemodel}d.
To investigate this in more detail, \cref{fig:dipolefull}  displays small frequency bandwidth spectra for various $k_\parallel$, in a similar presentation as previously discussed for the experiment. 
From this figure we directly observe a dip in $E$ (defined as $E\equiv 1-|r'|^2/|r_\text{glass}|^2$, see main text) similar to that we measured our experiment. 
Moreover, the depth of this dip significantly depends on the angle of incidence, showing that the we can reproduce the main feature of our experiment using our model. 
The benefit of doing these calculations is the access to the (corrected) polarizability $\alpha_\mathrm{lat}$ of the array.
\Cref{fig:dipolefull}b shows the imaginary part of $\alpha_\mathrm{lat}$ as we obtain it from our model.
Similar to the reflectance, this plot shows that the backaction and its effect on $\alpha_\mathrm{lat}$ can be controlled via the incident angle of the incoming drive field.
Interestingly, whereas the specular reflectance (\cref{fig:dipolefull}a) displays a clear dip, the polarizability has a more Fano-like resonant signature. 
The polarizability of the array thus quickly changes around the cavity resonance frequency.
We attribute this somewhat surprising discrepancy in shape between $E$ and $\text{Im}[\alpha_\mathrm{lat}]$ to the fact that we do not perform a pure extinction measurement, but instead also partly probe $|\alpha_\mathrm{lat}|^2$, which can be related to scattering by the antennas.
The interplay between these two contributions, scattering and extinction, most likely gives rise to a more complex behaviour that results from the interplay between the real and imaginary parts of $\tens{G}_\mathrm{hom}$, $\tens{G}_\mathrm{refl}$ and $\tens{G}_\mathrm{c}$.
Finally, we point towards an observed difference in Fano lineshapes for $E$ as we calculate them in \cref{fig:dipolefull}a, and  observe them experimentally in Fig.~4b (main text).
For example, for large angle of incidence (bottom panel in both figures) the Fano lineshapes have opposite asymmetry, i.e. opposite phase. 
We attribute this difference to the positioning of the antennas.
In the experiment we place the array on the air side of the interface, while in the calculation we put them inside the glass environment.
As scatterers positioned on an interface also experience backaction that relates to reflections originating from the interface, the phase of these reflections (contained in the complex Fresnel coefficient) is important. 
For evanescent waves (the $(-2)$ diffraction order) incident from the air side, this phase is opposite to that experienced by evanescent waves  entering from the glass side.
This opposite phase changes the interference condition with the directly reflected (zero-order) light and alters the observed shape of the reflectance signal.

% -------------------------------------- 

\subsection{On the relation between Albedo, Purcell factor and cooperativity}
\label{sec:Albedo}

To conclude this theory section we  make a connection between the cooperativity, a normalized value that is used to quantify the coupling between two resonators in a general coupled oscillator model, and the Purcell factor of a single cavity mode.
To make this connection we compare \cref{eq:alpha_PurcellAnt} with the coupled oscillator model from the main text. 

To start, realize that is possible to rewrite \cref{eq:alpha_PurcellAnt} in terms of the Albedo $A$, defined as $A\equiv \gamma_\mathrm{r}/\gamma$, which is the ratio between the radiative losses (excluding radiation into the cavity) and total losses. 
Approximating $(\oa^2-\ooc^2)$ as $2\ooc(\oa-\ooc)$ we insert the Albedo and detuning $\Delta=\ooc-\oo_\mathrm{a}$ such that \cref{eq:alpha_PurcellAnt} gives
\begin{equation}
\aleff \overset{\oo=\oc}{=} \frac{-\beta / \ooc}{2\Delta + \im\gamma(1+AF)}.
\label{eq:alpha_albedo}
\end{equation}
From this expression we thus find that the effective response $\aleff$ of a single scatterer depends on the product of its Albedo and the Purcell factor of a cavity at the scatterer's position. 
More specifically, we can identify that the \textit{total} loss rate $\gamma$ of the scatterer is modified from its original value: its loss rate increases with a factor that is given by the $AF$ product. 
In \cref{sec:finitelattice} we showed that the single particle polarizability is identical in lineshape to the lattice polarizability for the parameters studied here~(\cref{Fig:SI-FiniteLattice}).
As a result, it is opportune to use \cref{eq:alpha_albedo} together with \cref{eq:rprime_Abajo} to predict the lineshape of a reflection signal coming from a particle array in the presence of a single cavity mode. 
 
\begin{figure}
\begin{center}
\includegraphics[]{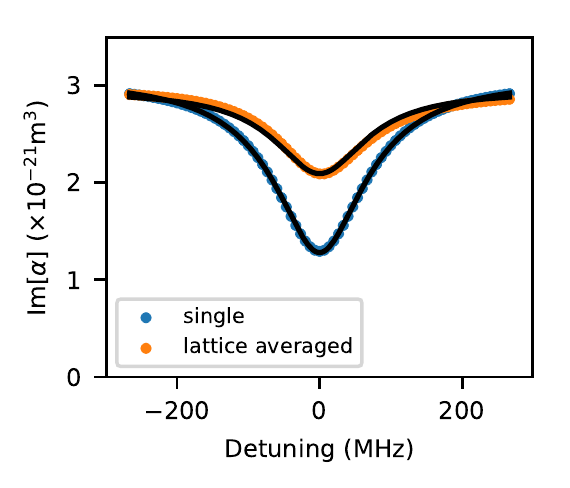} 
\caption{Polarizability spectrum of a single antenna coupled to the cavity (blue), and of a lattice of antennas (orange), where we average the antennas with a \SI{4.5}{\micro\meter} diameter real-space filter. 
Black lines show fits for $|\chi_\text{array}|$ (\cref{eq:coopbroadening}). Data is the same as shown in \cref{Fig:SI-FiniteLattice}.} 
\label{Fig:SI_fit_alphas} 
\end{center}
\end{figure}

To appreciate this result, we next want to identify a similar type of  response function and associated reflection signal, but retrieved from the coupled oscillator model introduced in the main text.
To achieve this, we reiterate the coupled oscillator model from the main text, starting at Eq.~(2). 
The array and cavity are resonators coupled at rate $g$ and are described by a Lorentzian response with complex field amplitudes $a$ and $c$, respectively.
We solve the driven system 
\begin{equation}
\begin{pmatrix}
\Delta_\mathrm{a} + \im \gamma /2 &  g \\
g & \Delta_\mathrm{c} + \im\kk/2
\end{pmatrix}
\begin{pmatrix}
a \\ c
\end{pmatrix}
= 
\begin{pmatrix}
\im \sqrt{\gamma_\mathrm{ex}} s_\mathrm{in} \\ 0
\end{pmatrix}
\end{equation} 
for $a$, and assume that both resonators are linear in frequency. 
Here we defined $\Delta_\mathrm{a}\equiv\omega-\oa$ and $\Delta_\mathrm{c}\equiv\omega-\oc$, where $\oo$ is the frequency of the incident field $s_\mathrm{in}$ driving the array and $\gamma_\mathrm{ex}$ the rate at which the array and input/output channel are coupled. 
Together with the input-output relation $s_\mathrm{out}= s_\mathrm{in} - \sqrt{\gamma_\mathrm{ex}} a$  (such that the reflection $r'=s_\mathrm{out}/s_\mathrm{in}$) this results in 
\begin{equation}
\frac{s_\mathrm{out}}{s_\mathrm{in}}=
1 - \frac{\im\gamma_\mathrm{ex}}{\Delta_\mathrm{a} + \im\gamma/2 - \frac{g^2}{\Delta_\mathrm{c} + \im\kk/2}}.
\end{equation}
From here, we proceed by parametrizing the interaction between both resonators using the cooperativity $C$, defined as $C=4g^2/(\gamma\kappa)$ with $\gamma(\kappa)$ the total loss rate of the antennas(cavity). 
This yields
\begin{equation}
\frac{s_\mathrm{out}}{s_\mathrm{in}}=
1 - \frac{2\im\gamma_\text{ex}} {2\Delta_\mathrm{a} + \im\gamma(1  
 + \frac{C}{\frac{2\Delta_\mathrm{c}}{\im\kappa} + 1})},
 \label{eq:alpha_coopbroadening}
\end{equation}
It is important to note that the second term on the right hand side of \cref{eq:alpha_coopbroadening} (the fraction) is associated with the response of the array, and in that sense should be compared to \cref{eq:alpha_albedo}.
Assuming that the incident field is resonant with the cavity ($\oo=\oc$, such that $\Delta_\mathrm{c}= 0$ and $\Delta_\text{a}=\Delta$), the response of the array $\chi_\text{array}$ as a function of cooperativity reads
\begin{equation}
\chi_\text{array} \overset{\oo=\oc}{=} \frac{- 2\im\gamma_\text{ex}}{2\Delta + \im \gamma(1 + C)}
\label{eq:coopbroadening}.
\end{equation}
Comparing \cref{eq:alpha_albedo,eq:coopbroadening} we learn that the Albedo Purcell factor product, the $AF$ product, is equivalent to the cooperativity $C$ in a more general two couped oscillator model: Both the $AF$ product and $C$ increase the total loss rate $\gamma$. As the Albedo by definition is always $\leq 1$, the cooperativity thus provides us with a lower bound on the Purcell factor. 
In the experiment we measured an \textit{average} cooperativity of antennas in the lattice, as more than one antenna was located within our real-space filter. In \cref{sec:finitelattice}, we have seen that we can relate the average antenna response to that of a single antenna. Thus, we can apply our knowledge of the lattice response to make an estimate for the cooperativity felt by a single antenna. \Cref{Fig:SI_fit_alphas} shows fits of the single particle and averaged lattice polarizability from \cref{Fig:SI-FiniteLattice} (c) with \cref{eq:coopbroadening}, from which we can extract a cooperativity for both responses. We find a cooperativity of 1.4 and 0.41 for the single particle and the averaged lattice response, respectively. The former is in good agreement with the $AF$ product in these calculations, i.e $A=0.15$, $F=9.1$, $AF=1.37$. We therefore expect that the experimentally measured maximum cooperativity of 0.5 implies a cooperativity of 1.7 for a single particle (in absence of other scatterers) located at the lattice origin. Thus the Purcell factor at the lattice origin must be higher than 1.7.

%\bibliography{alpha_biblio}

%

\end{document}